\documentclass[aps,prl,twocolumn,superscriptaddress,floatfix,showpacs,groupedaddress]{revtex4-1}

\usepackage{hyperref}
\usepackage{graphicx}
\usepackage{amsfonts}
\usepackage{amssymb}
\usepackage{amsmath}
\usepackage{txfonts}
\usepackage{lipsum}
\usepackage{color}
\usepackage{wasysym}

\usepackage{epstopdf}

\newcommand{\av}[1]{\ensuremath{\left\langle #1 \right\rangle}}
\newcommand{\qv}{\mathbf{q}}

\usepackage{letltxmacro}
\LetLtxMacro{\oldsqrt}{\sqrt}
\renewcommand{\sqrt}[2][\mkern8mu]{\mkern-6mu\mathop{}\oldsqrt[#1]{#2}}

\begin{document}

\title{Heisenberg-exchange-free nanoskyrmion mosaic}

\author{E. A. Stepanov$^{1,2}$, S. A. Nikolaev$^{2}$, C. Dutreix$^{3}$, M. I. Katsnelson$^{1,2}$, V. V. Mazurenko$^{2}$}
\affiliation{
$^1$Radboud University, Institute for Molecules and Materials, Nijmegen, The Netherlands\\
$^2$Department of Theoretical Physics and Applied Mathematics, Ural Federal University, Mira Street 19, 620002 Ekaterinburg, Russia\\
$^3$Univ Lyon, Ens de Lyon, Univ Claude Bernard, CNRS, Laboratoire de Physique, Lyon, France
}

\begin{abstract}
Isotropic Heisenberg exchange naturally appears as the main interaction in magnetism, usually favouring long-range spin-ordered phases. The anisotropic Dzyaloshinskii-Moriya interaction arises from relativistic corrections and is \textit{a priori} much weaker, even though it may sufficiently compete with the isotropic one to yield new spin textures. Here, we challenge this well-established paradigm, and propose to explore a Heisenberg-exchange-free magnetic world. There, the Dzyaloshinskii-Moriya interaction induces magnetic frustration in two dimensions, from which the competition with an external magnetic field results in a new mechanism producing skyrmions of nanoscale size. The isolated nanoskyrmion can already be stabilized in a few-atom cluster, and may then be used as LEGO${\textregistered}$ block to build a large magnetic mosaic. The realization of such topological spin nanotextures in $sp$- and $p$-electron compounds or in ultracold atomic gases would open a new route toward robust and compact magnetic memories.
\end{abstract}

\maketitle 

The concept of spin was introduced by G. Uhlenbeck and S. Goudsmit in the 1920s in order to explain the emission spectrum of the hydrogen atom obtained by A. Sommerfeld~\cite{Int2}. W.~Heitler and F.~London subsequently realized that the covalent bond of the hydrogen molecule involves two electrons of opposite spins, as a result of the fermionic exchange~\cite{Int4}. This finding inspired W. Heisenberg to give an empirical description of ferromagnetism~\cite{Int5,Int6}, before P. Dirac finally proposed a Hamiltonian description in terms of scalar products of spin operators~\cite{Int7}. These pioneering works focused on the ferromagnetic exchange interaction that is realized through the direct overlap of two neighbouring electronic orbitals. Nonetheless, P. Anderson understood that the exchange interaction in transition metal oxides could also rely on an indirect antiferromagnetic coupling via intermediate orbitals~\cite{PhysRev.115.2}. This so-called superexchange interaction, however, could not explain the weak ferromagnetism of some antiferromagnets. The latter has been found to arise from anisotropic interactions of much weaker strength, as addressed by I. Dzyaloshinskii and T. Moriya~\cite{DZYALOSHINSKY1958241,Moriya}. The competition between the isotropic exchange and anisotropic Dzyaloshinskii-Moriya interactions (DMI) leads to the formation of topologically protected magnetic phases, such as skyrmions~\cite{NagaosaReview}. Nevertheless, the isotropic exchange mainly rules the competition, which only allows the formation of large magnetic structures, more difficult to stabilize and manipulate in experiments~\cite{NagaosaReview,PhysRevX.4.031045}. Finding a new route toward more compact robust spin textures then appears as a natural challenge.

As a promising direction, we investigate the existence of two-dimensional skyrmions in the absence of isotropic Heisenberg exchange. Indeed, recent theoretical works have revealed that antiferromagnetic superexchange may be compensated by strong ferromagnetic direct exchange interactions at the surfaces of $sp$- and $p$-electron nanostructures~\cite{silicon, graphene}, whose experimental isolation has recently been achieved~\cite{PbSn,PhysRevLett.98.126401, SurfMagn, kashtiban2014atomically}. Moreover, Floquet engineering in such compounds also offers the possibility to dynamically switch off the isotropic Heisenberg exchange interaction under high-frequency-light irradiation, a unique situation that could not be met in transition metal oxides in equilibrium~\cite{PhysRevLett.115.075301,Control1,Control2}. In particular, rapidly driving the strongly-correlated electrons may be used to tune the magnetic interactions, which can be described by in terms of spin operators $\hat{\bf S}_i$ by the following Hamiltonian
\begin{align}
H_{\rm spin} = -\sum_{\av{ij}} J_{ij} (A) \,\hat{\bf S}_{i}\,\hat{\bf S}_{j} +
\sum_{\av{ij}}{\bf D}_{ij} (A)\,[\hat{\bf S}_{i}\times\hat{\bf S}_{j}],
\label{Hspin}
\end{align}
where the strengths of isotropic Heisenberg exchange $J_{ij}(A)$ and anisotropic DMI ${\bf D}_{ij}(A)$ now depend on the light amplitude $A$. The summations are assumed to run over all nearest-neighbour sites $i$ and $j$. The isotropic Heisenberg exchange term describes a competition between ferromagnetic direct exchange and antiferromagnetic kinetic exchange~\cite{PhysRev.115.2}. Importantly, it may be switched off dynamically by varying the intensity of the high-frequency light, while the anisotropic DMI remains non-zero~\cite{Control2}. 
\begin{figure}[b!]
\begin{center}
\includegraphics[width=0.67\linewidth]{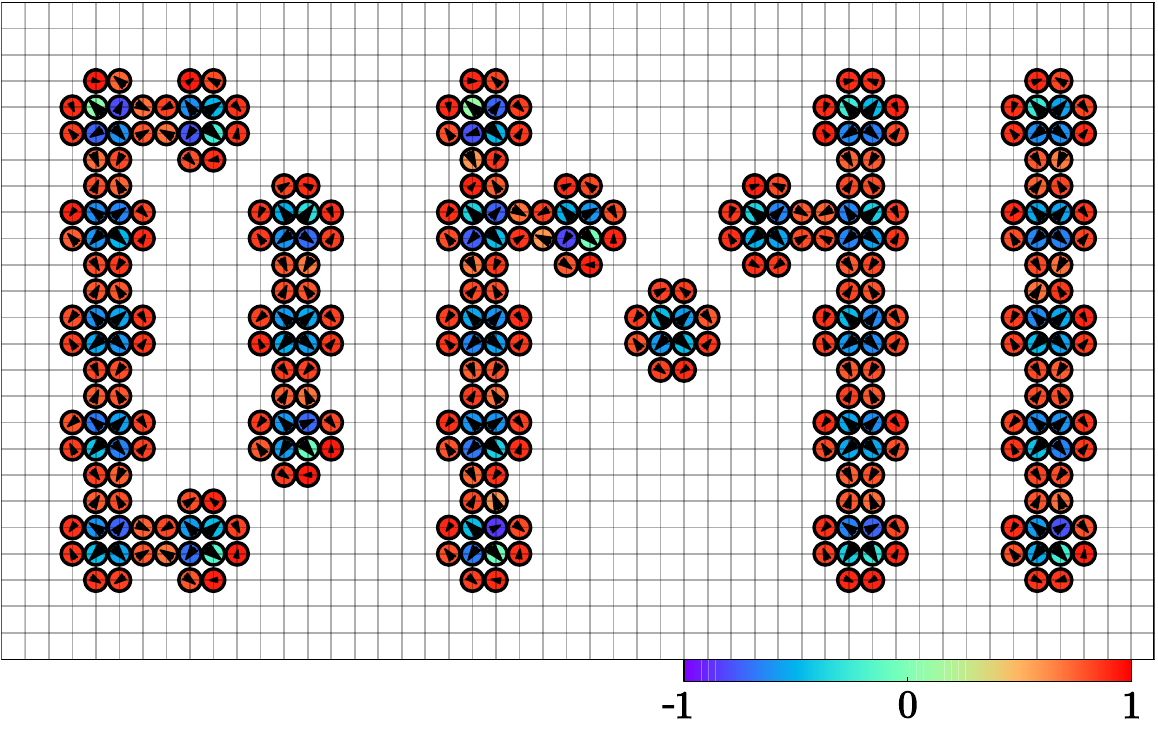}
\caption{Stable nanoskyrmion-designed DMI abbreviation resulting from the Monte Carlo simulation of the Heisenberg-exchange-free model on the non-regular square lattice with $B_{z} = 1.2$. Arrows and color depict the in- and out-of-plane spin projection, respectively.}
\label{Fig1}
\end{center}
\end{figure}
The study of Heisenberg-exchange-free magnetism may also be achieved in other classes of systems, such as optical lattices of ultracold atomic gases. Indeed, cold atoms have enabled the observation and control of superexchange interactions, which could be reversed between ferromagnetic and antiferromagnetic~\cite{Trotzky}, as well as strong DMI~\cite{Gong}, following the realization of the spin-orbit coupling in bosonic and fermionic gases~\cite{SO_Lin,SO_Wang}.

Here, we show that such a control of the microscopic magnetic interactions offers an unprecedented opportunity to observe and manipulate nano-scale skyrmions. Heisenberg-exchange-free nanoskyrmions actually arise from the competition between anisotropic DMI and a constant magnetic field. The latter was essentially known to stabilize the spin textures~\cite{PhysRevX.4.031045}, whereas here it is a part of the substantially different and unexplored mechanism responsible for nanoskyrmions. Fig.~\ref{Fig1} immediately highlights that an arbitrary system of few-atom skyrmions can be stabilized and controlled on a non-regular lattice with open boundary conditions, which was not possible at all in the presence of isotropic Heisenberg exchange.

{\it Heisenberg-exchange-free Hamiltonian} ---
Motivated by the recent predictions and experiments discussed above, we consider the following spin Hamiltonian
\begin{align}
&\hat H_{\rm Hef} = 
\sum_{\av{ij}}{\bf D}_{ij}\,[\hat{\bf S}_{i}\times\hat{\bf S}_{j}] - \sum_{i}{\bf B}\hat{\bf S}_{i},
\label{DMI}
\end{align}
where the magnetic field is perpendicular to the two dimensional system and ${\bf B}=~(0,0,B_{z})$. The latter tends to align the spins in the $z$ direction, while DMI flavors their orthogonal orientations. At the quantum level this non-trivial competition provides a fundamental resource in quantum information processing~\cite{QuantInf}. Here, we are interested in the semi-classical description of the Heisenberg-exchange-free magnetism.
\begin{figure}[t!]
\begin{center}
\includegraphics[width=1\linewidth]{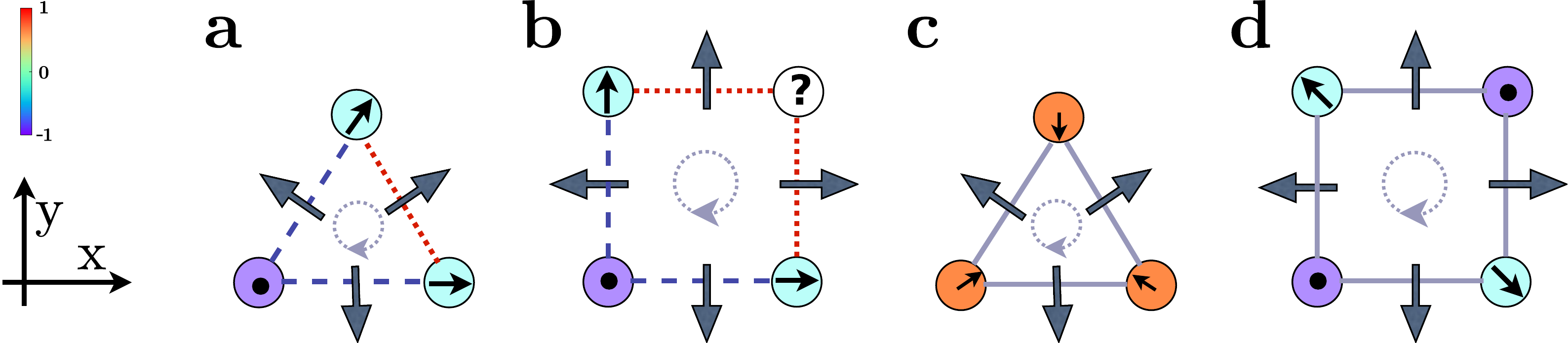}
\caption{Magnetic frustration in elementary DMI clusters of the triangular ({\bf a}) and square ({\bf b}) lattices. Curving arrows denote clockwise direction of the bonds in each cluster. Big gray arrows correspond to the in-plane DMI vectors. Black arrows in circles denote the in-plane directions of the spin moments.  Blue dashed and red dotted lines indicate the bonds with minimal and zero DMI energy, respectively. ({\bf c}) and ({\bf d}) illustrate the examples of the spin configurations corresponding to classical ground state of the DMI Hamiltonian.}
\label{Fig2}
\end{center}
\end{figure}
{\it DMI-induced frustration} ---
In the case of the two-dimensional materials with pure DMI between nearest neighbours, magnetic frustration is the intrinsic property of the system. To show this let us start off with the elementary plaquettes of the triangular and square lattices without external magnetic field (see Fig.~\ref{Fig2}). Keeping in mind real two-dimensional materials with the $C_{nv}$ symmetry~\cite{graphene,silicon} we consider the in-plane orientation of the DMI vector perpendicular to the corresponding bond. Taking three spins of a single square plaquette as shown in Fig.~\ref{Fig2}~{\bf b}, one can minimize their energy while discarding their coupling with the fourth spin. Then, the orientation of the remaining spin can not be uniquely defined, because the spin configuration regardless whether it points ``up'' or ``down'' has the same energy, which indicates frustration. Thus, Fig.~\ref{Fig2}~{\bf d} gives the example of the classical ground state of the square plaquette with the energy ${\rm E}_{\square} = - \sqrt{2}\,|\mathbf{D}_{ij}|\,S^2$ and magnetization ${\rm M}^z_{\square} =  S/2$ (per spin). In turn, frustration of the triangular plaquette is expressed in Fig.~\ref{Fig2}~{\bf b}, while its magnetic ground state is characterized by the following spin configuration 
$\mathbf{S}_1 = (0, -\frac{1}{\sqrt{2}}, \frac{1}{\sqrt{2}})\,S$,  $\mathbf{S}_2 = (\frac{\sqrt{3}}{2\sqrt{2}}, \frac{1}{2\sqrt{2}}, \frac{1}{\sqrt{2}})\,S$ and $\mathbf{S}_3 = (-\frac{\sqrt{3}}{2\sqrt{2}}, \frac{1}{2\sqrt{2}}, \frac{1}{\sqrt{2}})\,S$ shown in Fig.~\ref{Fig2}~{\bf c}. One can see that the in-plane spin components form $120^{\circ}$-Neel state similar to the isotropic Heisenberg model on the triangular lattice~\cite{PhysRev.115.2}. The corresponding energy and magnetization are ${\rm E}_{\triangle} =- \sqrt{3}\,|\mathbf{D}_{ij}|\,S^2$ and ${\rm M}^z_{\triangle} = S/\sqrt{2}$, respectively. 

Importantly, the ground state of the triangular and square plaquettes is degenerate due to the $C_{nv}$ and in-plane mirror symmetry. For instance, there is another state with the same energy ${\rm E'} = {\rm E}$, but opposite magnetization ${\rm M}'^z = - {\rm M}^z$.
Therefore, such elementary magnetic units can be considered as the building blocks in order to realize {\it spin spirals}
on lattices with periodic boundary conditions and zero magnetic field. The ensuing results are obtained via Monte Carlo simulations, as detailed in Supplemental Material~\cite{SM}.

\begin{figure}[t!]
\begin{center}
\includegraphics[width=1\linewidth]{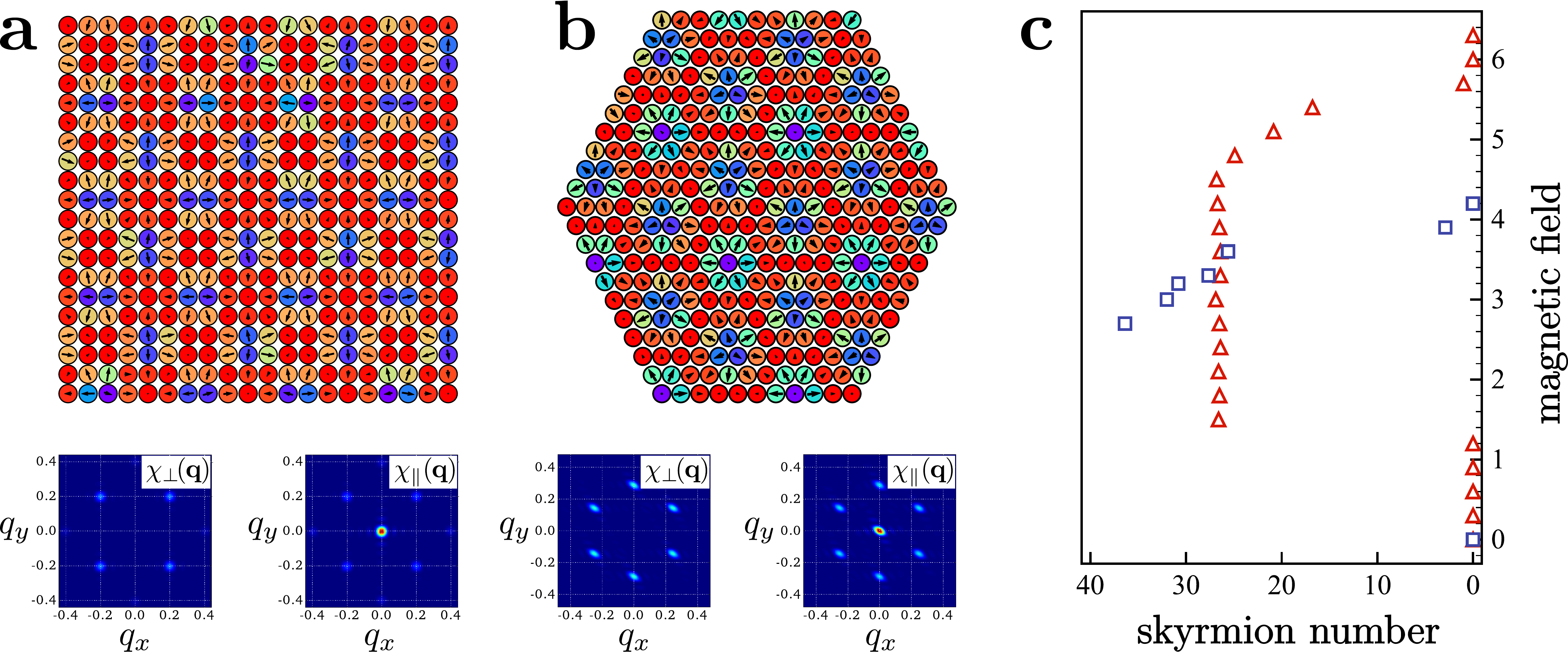}
\caption{Fragments of the spin textures and spin structure factors obtained with the Heisenberg-exchange-free model on the square $20\times20$ ({\bf a}) and triangular $21\times21$ ({\bf b}) lattices. The values of the magnetic fields in these simulations were chosen $B_{z} = 3.0$ and $B_{z} = 3.2$  for the triangular and square lattices, respectively.  The calculated skyrmion numbers for the triangular (blue triangles) and square (red squares) lattices ({\bf c}). The magnetic field is in units of DMI. The temperature is equal to ${\rm T}=0.01\,|{\rm \bf D}|$.}
\label{Fig3}
\end{center}
\end{figure}

{\it Nanoskyrmionic state} --- 
At finite magnetic field the spiral state can be transformed to a skyrmionic spin texture. Fig.~\ref{Fig3} gives some examples obtained from the Monte Carlo calculations for the triangular and square lattices. Remarkably, the radius of the obtained nanoskyrmions does not exceed a few lattice constants. The calculated spin structure factors $\chi_{\perp}(\qv)$ and $\chi_{\parallel}(\qv)$~\cite{SM} revealed a superposition of the two (square lattice) or three (triangular lattice) spin spirals with $\pm \mathbf{q}$, which is the first indication of the skyrmionic state (see Fig.~\ref{Fig3}~{\bf a}, {\bf b}). For another confirmation one can calculate the skyrmionic number, which is related to the topological charge. In the discrete case, the result is extremely sensitive to the number of sites comprising a single skyrmion and to the way the spherical surface is approximated~\cite{Rosales}.  Here we used the approach of Berg and L\"uscher~\cite{Berg}, which is based on the definition of the topological charge as the sum of the nonoverlapping spherical triangles areas~\cite{Blugel2016,SM}. According to our simulations, the topological charge of each object shown in Fig.~\ref{Fig3}~{\bf a} and Fig.~\ref{Fig3}~{\bf b} is equal to unity. The square and triangular lattice systems exhibit completely different dependence of the average skyrmion number on the magnetic field. As it is shown in Fig.~\ref{Fig3}~{\bf c}, the skyrmionic phase for the square lattice $2.7\leq B_{z} < 4.2$ is much narrower than the one of the triangular lattice 
$1.2 \leq B_{z} < 6$. Moreover, in the case of the square lattice we observe strong finite-size effects leading to a non-stable value of the average topological charge in the region of $0 < B_{z} < 2.7$. Since the value of the magnetic field is given in the units of DMI, the topological spin structures in the considered model require weak magnetic fields, which is very appealing for modern experiments. 

We would like to stress that the underlying mechanism responsible for the skyrmions presented in this work is intrinsically different from those presented in other studies. Generally, skyrmions can be realized by means of the different mechanisms~\cite{NagaosaReview}. For instance, in noncentrosymmetric systems these spin textures arise from the competition between isotropic and anisotropic exchange interactions. On the other hand, magnetic frustration induced by competing isotropic exchange interactions can also lead to a skyrmion crystal state, even in the absence of DMI and anisotropy. Moreover, following the results of~\cite{Blugel} the nanoskyrmions are stabilized due to a four-spin interaction. Nevertheless, the nanoskyrmions have never been predicted and observed as the result of the interplay between DMI and a constant magnetic field.

\begin{figure}[t!]
\begin{center}
\includegraphics[width=0.64\linewidth]{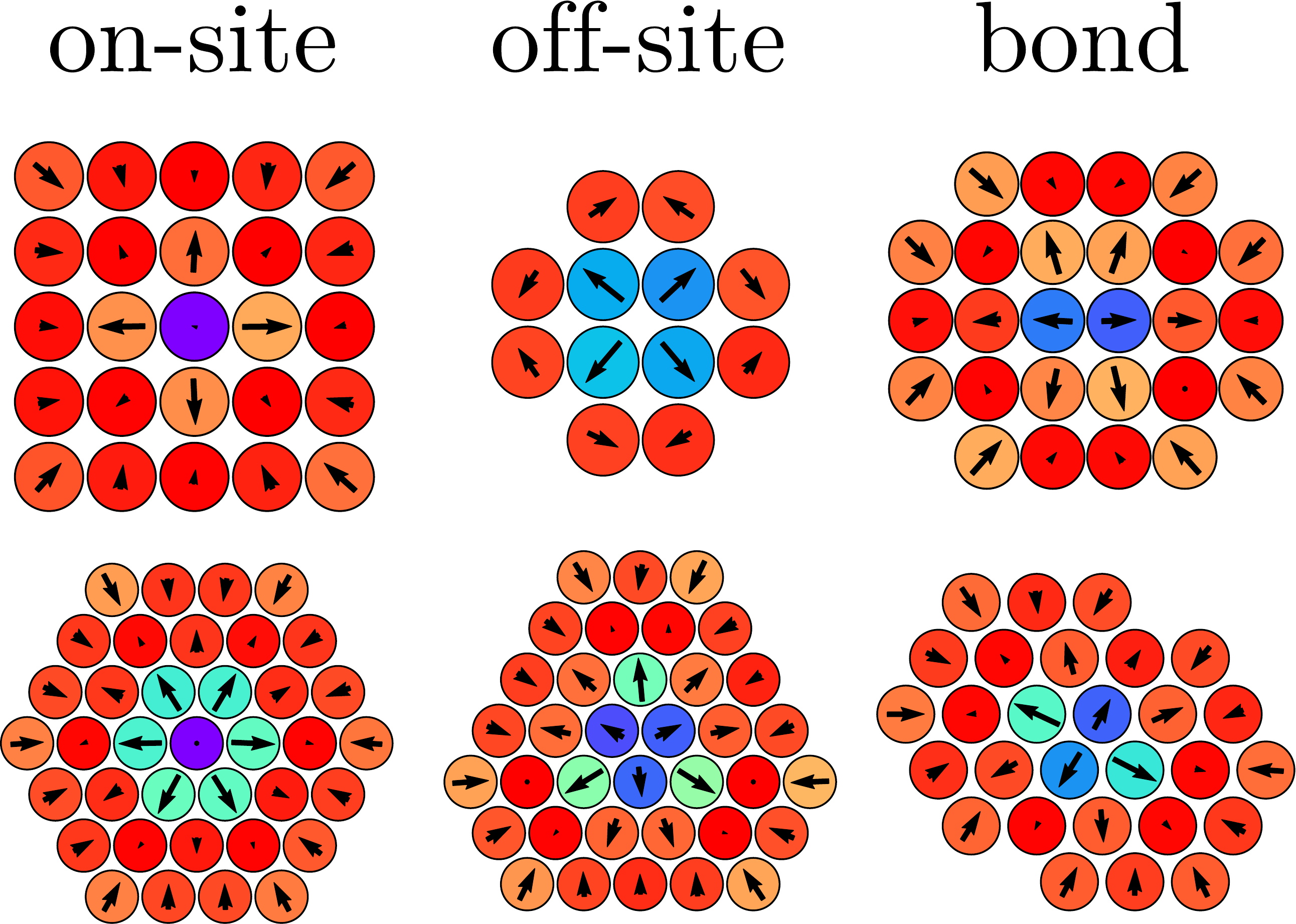}
\caption{Catalogue of the DMI nanoskyrmion species stabilized on the small square (top figures) and triangular (bottom figures) clusters with open boundary conditions. The corresponding magnetic fields are (from top to bottom): on-site $B_{z}=3.0;~3.0$, off-site $B_{z}=1.2;~2.4$, bond-centered $B_{z}=2.5;~3.0$. }
\label{Fig4}
\end{center}
\end{figure}
Full catalogue of nanoskyrmions obtained in this study is presented in Fig.~\ref{Fig4}. As one can see, they can be classified with respect to the position of the skyrmionic center on the discrete lattice. Thus, the on-site, off-site (center of an elementary triangle or square) and bond-centred configurations have been revealed. Importantly, these structures can be stabilized not only on the lattice, but also on the isolated plaquettes of 12-37 sites with open boundary conditions. It is worth mentioning that not all sites of the isolated plaquettes form the skyrmion. In some cases inclusion of the additional spins that describe the environment of the skyrmion is necessary for stabilization of the topological spin structure. As we discuss below, it can be used to construct a magnetic domain for data storage.

\begin{figure}[t!]
\begin{center}
\includegraphics[width=1\linewidth]{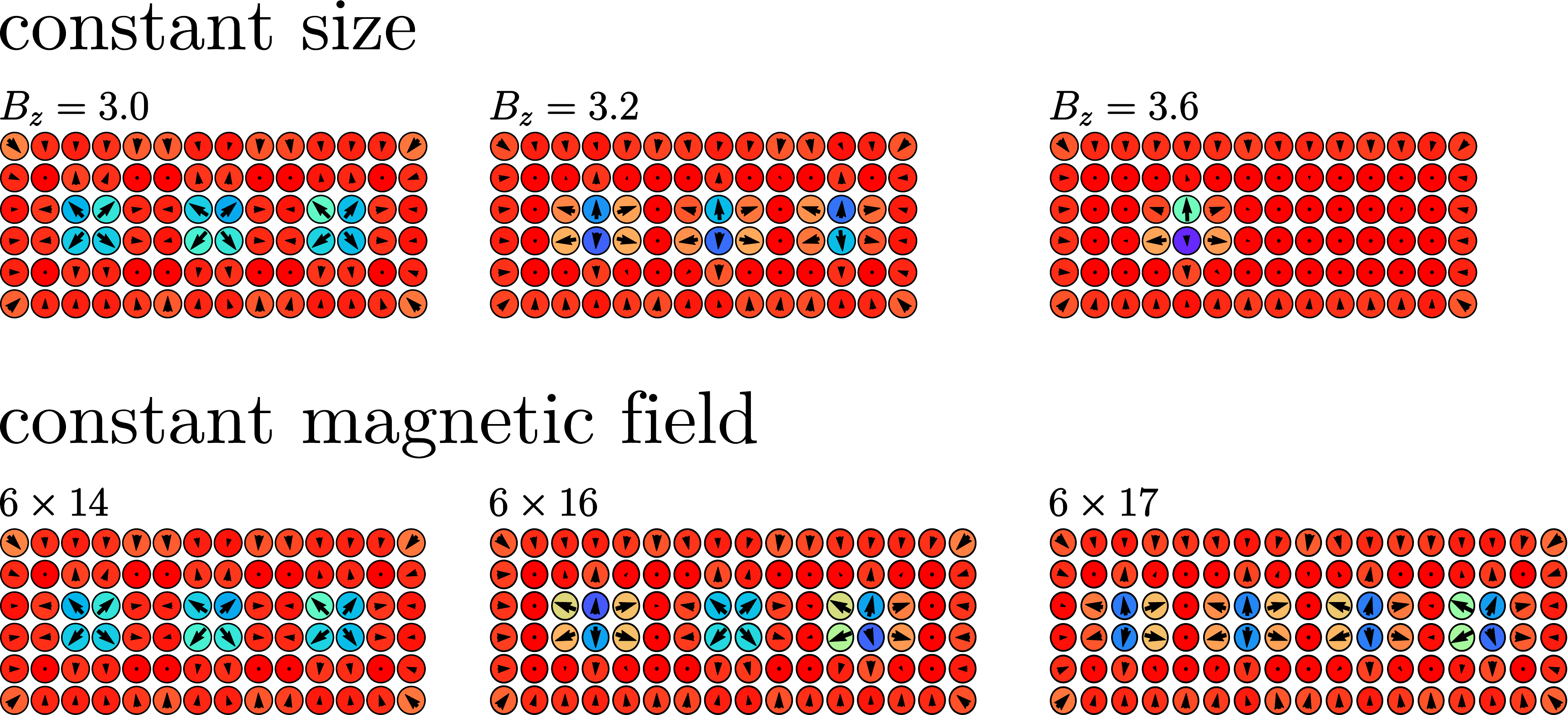}
\caption{(Top panel) Evolution of the nanoskyrmions on the lattice $6\times14$ with respect to the magnetic field. (Bottom panel) Examples of nanoskyrmion mosaics obtained from the Monte Carlo simulations of the DMI model on the square lattices with open boundary conditions at the magnetic field $B_{z}=3$.}
\label{Fig5}
\end{center}
\end{figure}

{\it Nanoskyrmionic mosaic} ---  
By defining the local rules (interaction between nearest neighbours) and external parameters (such as the lattice size and magnetic field) one can obtain magnetic structures with different patterns by using Monte Carlo approach. As it follows from Fig.~\ref{Fig5}, the pure off-site square skyrmion structures are realized on the lattices $6\times14$ with open boundary conditions at $B_{z} = 3.0$. Increasing the lattice size along the $x$ direction injects bond-centred skyrmions into the system. In turn, increasing the magnetic field leads to the compression of the nanoskyrmions and reduces their density.  At $B_{z}=3.6$ we observed the on-site square skyrmion of the most compact size. Thus the particular pattern of the resulting nanoskyrmion mosaic respect the minimal size of individual  nanoskyrmions and the tendency of the system to formation of close-packed structures to minimize the energy.

The solution of the Heisenberg-exchange-free Hamiltonian~\eqref{DMI} for the magnetic fields corresponding to the skyrmionic phase can be related to  the famous geometrical NP-hard problem of bin packing~\cite{Bin}. Let us imagine that there is a set of fixed-size and fixed-energy objects (nanoskyrmions) that should be packed on a lattice of $n \times m$ size in a more compact way. As one can see, such objects are weakly-coupled to each other. Indeed, the contact area of different skyrmions is nearly ferromagnetic, so the binding energy between two skyrmions is very small, since it is related to DMI. In addition, the energy difference between nanoskyrmions of different types is very small as can be seen from Fig.~\ref{Fig5}. Indeed, the energy difference between three off-site and bond-centered skyrmions realized on the $6\times14$ plaquette is about $0.2\,B_{z}$.  Here, we sample spin orientation within the Monte Carlo simulations and do not manipulate the nanoskyrmions directly. Thus, the stabilization of a periodic and close-packed nanoskyrmionic structures on the square or triangular lattice with the length of more than $30$ sites becomes a challenging task. Therefore, the problem can be addressed to a LEGO$\textregistered$-type constructor, where one builds the mosaic pattern using the unit nanoskyrmionic bricks. 

{\it Size limit} --- 
For practical applications, it is of crucial importance to have skyrmions with the size of the nanometer range, for instance to achieve a high density memory~\cite{Lin}. Previously the record density of the skyrmions was reported for the Fe/Ir(111) system~\cite{Blugel} for which the size of the unit cell of the skyrmion lattice stabilized due to a four-spin interaction was found to be 1nm $\times$ 1nm.  In our case the diameter of the triangular on-site skyrmion is found to be $4$ lattice constants (Fig.~\ref{Fig4}). Thus for prototype $sp$-electron materials the diameter is equal to 1.02 nm (semifluorinated graphene~\cite{graphene}) and 2.64 nm (Si(111):\{Sn,Pb\}~\cite{silicon}). 

On the basis of the obtained results we predict the smallest diameter of $2$ lattice constants for the on-site square skyrmion. Our simulations for finite-sized systems with open boundary conditions show that such a nanoskyrmion can exist on the $5\times5$ cluster (Fig.~\ref{Fig4} top right plaquette), which is smaller than that previously reported in~\cite{Keesman}. We believe that this is the ultimate limit of a skyrmion size on the square lattice.

{\it Micromagnetic model} --- 
The analysis of the isolated skyrmion can also be fulfilled on the level of the micromagnetic model treating the magnetization as a continuous vector field~\cite{SM}. Contrary to the case of nonzero exchange interaction, the Heisenberg-exchange-free Hamiltonian~\eqref{DMI} allows to obtain an analytical solution for the skyrmionic profile. In the particular case of the square lattice, the radius of the isolated skyrmion is equal to $R=4Da/B$, where $a$ is the lattice constant. Moreover, the skyrmionic solution is stable even in the presence of a small exchange interaction $J\ll{}D$~\cite{SM}. 
It is worth mentioning that the obtained result for the radius of the Heisenberg-exchange-free skyrmion is essentially different from the case of competing exchange interaction and DMI, where the radius is proportional to the ratio $J/D$. Although in the absence of DMI both, the exchange interaction and magnetic field, favour the collinear orientation of spins in the direction perpendicular to the surface, the presence of DMI changes the picture drastically. When the spins are tilted by the anisotropic interaction,  the magnetic field still wants them to point in the $z$ direction, while the exchange interaction tries to keep two neighbouring spins parallel without any relation to the axes. Therefore, the stronger magnetic field decreases the radius of the skyrmion, while the larger value of exchange interaction broadens the structure~\cite{ref}. 

\begin{figure}[b!]
\begin{center}
\includegraphics[width=0.9\linewidth]{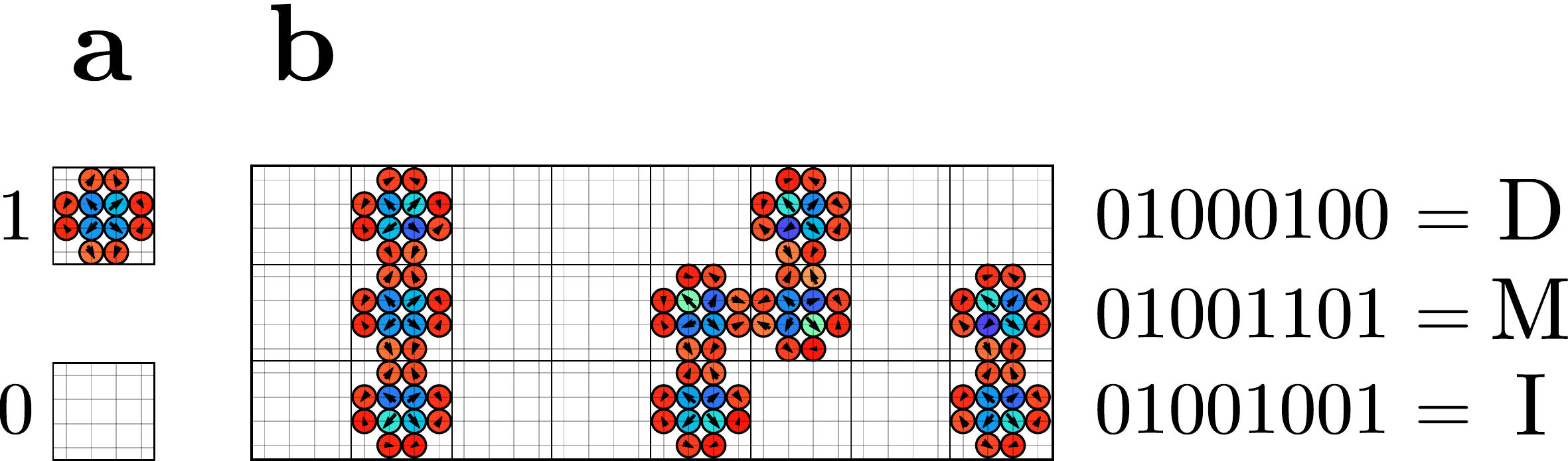}
\caption{({\bf a}) two possible states of the nanoskyrmionic bit. ({\bf b}) the 24-bit nanoskyrmion memory block encoding DMI abbreviation as obtained from the Monte Carlo simulations with $B_{z} =1.2$.}
\label{Fig6}
\end{center}
\end{figure}

{\it Memory prototype} --- 
Having analyzed individual nanoskyrmions we
are now in a position to discuss technological applications of the
nanoskyrmion mosaic.
Fig.~\ref{Fig6} a visualizes a spin structure consisting of the
elementary blocks of two types that we associated with the two
possible states of a single bit, the ``1'' and ``0''. According to our
Monte Carlo simulations a side stacking of off-site square plaquette
visualized in Fig.~\ref{Fig4} protects skyrmionic state in each
plaquette. Thus we have a stable building block for design of the
nano-scale memory or nanostructures presented in Fig.~\ref{Fig1}.
Similar to the experimentally realized vacancy-based memory~\cite{Memory}, a specific filling of the lattice can be reached
by means of the scanning tunnelling microscopy (STM) technique. In
turn, the spin-polarized regime~\cite{STM} of STM is
to be used to read the nanoskyrmionic state. The density of the memory
prototype we discuss can be estimated as 1/9 bits $a^{-2}$ ($a$ is the
lattice constant), which is of the same order of magnitude as obtained
for vacancy-based memory.

{\it Conclusion} --- We have introduced a new class of the two-dimensional systems that are described with the Heisenberg-exchange-free Hamiltonian. The frustration of DMI on the triangular and square lattices leads to a non-trivial state of nanoskyrmionic mosaic that can be manipulated by varying the strength of the constant magnetic field and the size of the sample. Importantly, such a state appears as a result of competition between DMI and the constant magnetic field. This mechanism is unique and is reported for the first time. Being stable on non-regular lattices with open boundary conditions, nanoskyrmionic phase is shown to be promising for technological applications as a memory component. We also present the catalogue of nanoskyrmionic species that can be stabilized already on a tiny plaquettes of a few lattice sites. Characteristics of the isolated skyrmion were studied both, numerically and analytically, within the Monte Carlo simulations and in the framework of the micromagnetic model, respectively.

\begin{acknowledgments}
We thank Frederic Mila, Alexander Tsirlin and Alexey Kimel for fruitful discussions. The work of E.A.S. and V.V.M. was supported by the Russian Science Foundation, Grant 17-72-20041. The work of M.I.K. was supported by NWO via Spinoza Prize and by ERC Advanced Grant 338957 FEMTO/NANO. Also, the work was partially supported by the Stichting voor Fundamenteel Onderzoek der Materie (FOM), which is financially supported by the Nederlandse Organisatie voor Wetenschappelijk Onderzoek (NWO).
\end{acknowledgments}

\clearpage
\onecolumngrid

\begin{center}
\Large{Supplemental Material for ``Heisenberg-exchange-free nanoskyrmion mosaic''}
\end{center}

\section{Methods}
The DMI Hamiltonian with classical spins was solved by means of the
Monte Carlo approach. The spin update scheme is based on the
Metropolis algorithm. The systems in question are gradually (200
temperature steps) cooled down from high temperatures (${\rm T}\sim
|\mathbf{D}_{ij}|$) to ${\rm T}=~0.01|\mathbf{D}_{ij}|$. Each temperature
step run consists of $1.5\times10^{6}$ Monte Carlo steps. The
corresponding micromagnetic model was solved analytically.

\section{Definition of the skyrmion number}
Skyrmionic number is related to the topological
charge. In the discrete case, the result is extremely sensitive
to the number of sites comprising a single Skyrmion and to the
way the spherical surface is approximated. Here we
used the approach of Berg and L\"uscher, which is based
on the definition of the topological charge as the sum of the
nonoverlapping spherical triangles areas. Solid angle subtended by the spins ${\bf S}_{1}$, ${\bf S}_{2}$ and ${\bf S}_{3}$ is defined as
\begin{align}
A = 2 \arccos[\frac{1+ {\bf S}_{1} {\bf S}_{2} + {\bf S}_{2} {\bf S}_{3} + {\bf S}_{3} {\bf S}_{1}}{\sqrt{2(1+ {\bf S}_{1} {\bf S}_{2} )(1+ {\bf S}_{2} {\bf S}_{3})(1+ {\bf S}_{3} {\bf S}_{1})}}].
\end{align}
We do not consider the exceptional configurations for which
\begin{align}
&{\bf S}_{1} [{\bf S}_{2} \times {\bf S}_{1}] = 0 \\
&1+ {\bf S}_{1} {\bf S}_{2} + {\bf S}_{2} {\bf S}_{3} + {\bf S}_{3} {\bf S}_{1} \le 0. \notag
\end{align} 
Then the topological charge $Q$ is equal to
$
Q = \frac{1}{4\pi} \sum_{l} A_{l}.
$

\section{Spin spiral state}
\begin{figure}[b!]
\includegraphics[width=0.55\linewidth]{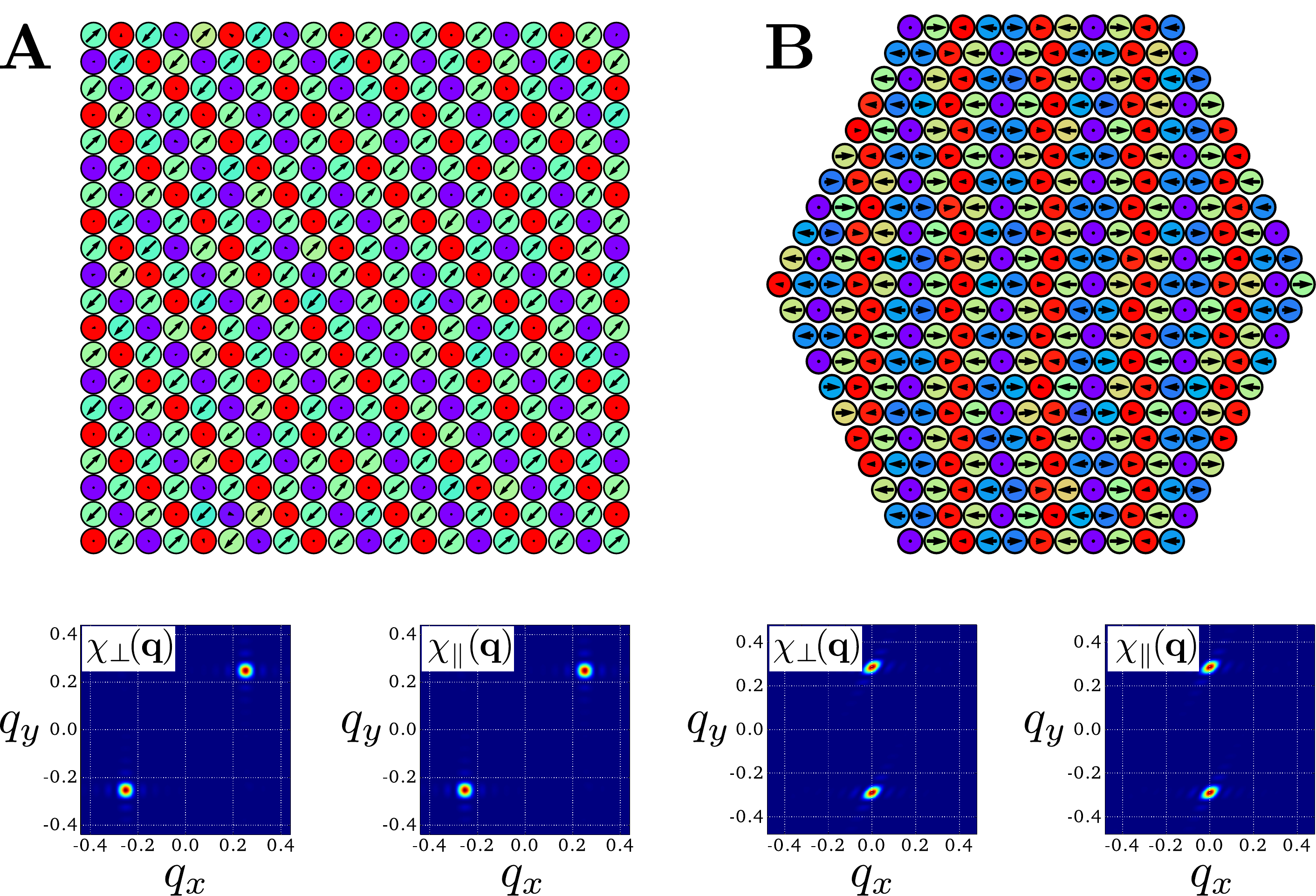}
\caption{Fragments of the spin textures and spin structure factors obtained with the Heisenberg-exchange-free model on the square $20\times20$ {\bf A} and triangular $21\times21$ {\bf B} lattices in the absence of the magnetic field. The temperature is equal to ${\rm T}=0.01\,|{\rm \bf D}|$.}
\label{spinfactors}
\end{figure}

Our Monte Carlo simulations for the DMI Hamiltonian
with classical spin $|\mathbf{S}| = 1$ have shown that the triangular and square lattice systems form {\it spin spiral} structures (see Fig.~\ref{spinfactors}).
The obtained spin textures and the calculated spin structure factors are
\begin{align}
\chi_{\perp}(\qv)&=\frac{1}{N}\av{\left|\sum_{i} S_{i}^{x} \, e^{-i\qv\cdot{\bf r}_{i}} \right|^{2}+\left|\sum_{i} S_{i}^{y} \, e^{-i\qv\cdot{\bf r}_{i}} \right|^{2}}\\
\chi_{\parallel}(\qv)&=\frac{1}{N}\av{\left|\sum_{i} S_{i}^{z} e^{-i\qv\cdot{\bf r}_{i}} \right|^{2}}.
\end{align}
The pictures of intensities at zero magnetic field correspond to the spin spiral state with $|\mathbf{q}_{\square}| = \frac{1}{2\sqrt{2}} \times \frac{2\pi}{a}$ and $|\mathbf{q}_{\triangle}| \simeq 0.29 \times \frac{2\pi}{a}$ for the square and triangle lattices, respectively. Here $a$ is the lattice constant. The corresponding periods of the spin spirals are $\lambda_{\triangle} = 3.5\,a$ and $\lambda_{\square} = 2 \sqrt{2}\,a$. The energies of the triangular and square systems at B$_{z}$ and low temperatures are scaled as energies of the elementary clusters, namely E$_{\triangle}$ and E$_{\square}$ (Fig.~\ref{Fig2}~c and Fig.~\ref{Fig2}~d), multiplied by the number of sites. In contrast to the previous considerations taking into account Heisenberg exchange interaction, we do not observe any long-wavelength spin excitations ($\qv=0$) in $\chi_{\parallel}(\boldsymbol{q})$.

\section{Micromagnetics of isolated skyrmion}

A qualitative description of a single skyrmion can be obtained within a micromagnetic model, when the quantum spin operators $\hat{\bf S}$ are replaced by the classical local magnetization ${\bf m}_{i}$ at every lattice site and later by a continuous and differentiable vector field ${\bf m}({\bf r})$ as $\hat{\bf S}_{i} \to S {\bf m}_{i} \to S{\bf m}({\bf r})$, where $S$ is the spin amplitude and $|{\bf m}|=1$. This approach is valid for large quantum spins when the length scale on which the magnetic structure varies is larger than the interatomic distance. Let us make the specified transformation explicitly and calculate the energy of the single spin localized at the lattice site $i$ that can be found as a sum of the initial DMI Hamiltonian over the nearest neighbor lattice sites
\begin{align}
\label{mme1}
E_{i}&=-\sum_{j}J_{ij}\,\hat{\bf S}_{i}\,\hat{\bf S}_{j} +
\sum_{j}{\bf D}_{ij}\,[\hat{\bf S}_{i}\times\hat{\bf S}_{j}] - 
{\bf B}\,\hat{\bf S}_{i} \\
&= -S^{2}\sum_{j}J_{ij}\,{\bf m}_{i}\,{\bf m}_{j} + 
S^2\sum_{j}{\bf D}_{ij}\,[{\bf m}_{i}\times{\bf m}_{j}] - 
S{\bf B}\,{\bf m}_{i} \notag\\
&=-S^{2}\sum_{j}J_{ij}\left(1-\frac{1}{2}({\bf m}_{j}-{\bf m}_{i})^{2}\right) + 
S^2\sum_{j}{\bf D}_{ij}\left[{\bf m}_{i}\times\left({\bf m}_{j}-{\bf m}_{i}\right)\right] - S{\bf B}\,{\bf m}_{i} \notag\\
&\simeq-S^{2}\sum_{j}J_{ij}\left(1-\frac{1}{2}({\bf m}({\bf r}_{i}+\delta{\bf  r}_{ij}) - {\bf m}({\bf r}_{i}))^{2}\right) + S^2\sum_{j}{\bf D}_{ij}\left[{\bf m}_{i}({\bf r}_{i})\times\left({\bf m}({\bf r}_{i}+\delta{\bf r}_{ij}) - {\bf m}({\bf r}_{i})\right)\right] - S{\bf B}\,{\bf m}({\bf r}_{i}) \notag\\
&\simeq\frac{S^{2}a^2}{2}\sum_{j}J_{ij}\left(({\bf e}_{ij}\nabla)\,{\bf m}({\bf r}_{i})\right)^{2} + S^2a\sum_{j}{\bf D}_{ij}\left[{\bf m}_{i}({\bf r}_{i})\times 
(({\bf e}_{ij}\nabla)\,{\bf m}({\bf r}_{i}))\right] - S{\bf B}\,{\bf m}({\bf r}_{i}), \notag
\end{align}
The DMI vector ${\bf D}_{ij} = D\left[{\bf e}_{z}\times{\bf e}_{ij}\right]$ is perpendicular to the vector ${\bf e}_{ij}$ that connects spins at the nearest-neighbor sites $\av{ij}$ and favours their orthogonal alignment, while the exchange term tends to make the magnetization uniform. The above derivation was obtained for a particular case of a square lattice, but can be straightforwardly generalized to an arbitrary configuration of spins. Then, we obtain
\begin{align}
\label{mme2}
E_{i} 
&= \frac{S^{2}a^2}{2}\sum_{j}J_{ij}\left(({\bf e}_{ij}\nabla)\,{\bf m}({\bf r}_{i})\right)^{2} + S^2aD\sum_{j}\left[{\bf e}_{z}\times{\bf e}_{ij}\right]\left[{\bf m}_{i}\times 
\Big(({\bf e}_{ij}\nabla)\,{\bf m}({\bf r}_{i})\Big)\right] - S{\bf B}_{z}\,{\bf m}_{z}({\bf r}_{i}) \\
&= \frac{J'}{2}\left[\Big(\partial_{x}\,{\bf m}({\bf r}_{i})\Big)^{2} + \Big(\partial_{y}\,{\bf m}({\bf r}_{i})\Big)^{2} \right] - S{\bf B}_{z}\,{\bf m}_{z}({\bf r}_{i}) \notag \\   
&\,+ D'\,\left[{\bf m}_{z}({\bf r}_{i})\,\Big(\partial_{x}{\bf m}_{x}({\bf r}_{i})\Big) - \Big(\partial_{x}{\bf m}_{z}({\bf r}_{i})\Big)\,{\bf m}_{x}({\bf r}_{i})) + {\bf m}_{z}({\bf r}_{i})\,\Big(\partial_{y}{\bf m}_{y}({\bf r}_{i})\Big) - \Big(\partial_{y}{\bf m}_{z}({\bf r}_{i})\Big)\,{\bf m}_{y}({\bf r}_{i}))\right] \notag
\end{align}
where $J'=2JS^{2}a^{2}$, $D'=2DS^{2}a$ and $B'=BS$. The unit vector of the magnetization at every point of the vector field can be parametrized by ${\bf m} = \sin\theta\cos\psi\,{\bf e}_{x} + \sin\theta\sin\psi\,{\bf e}_{y} + \cos\theta\,{\bf e}_{z}$ in the spherical coordinate basis. In order to describe axisymmetric skyrmions, we additionally introduce the cylindrical coordinates $\rho$ and $\varphi$, so that $\rho=0$ is associated to the center of a skyrmion
\begin{align}
\partial_{x}{\bf m}({\bf r}_{i}) &= 
\cos\varphi\,\partial_{\rho}{\bf m}({\bf r}_{i}) - 
\frac{1}{\rho}\sin\varphi\,\partial_{\varphi}{\bf m}({\bf r}_{i})\,, \\
\partial_{y}{\bf m}({\bf r}_{i}) &= 
\sin\varphi\,\partial_{\rho}{\bf m}({\bf r}_{i}) + 
\frac{1}{\rho}\cos\varphi\,\partial_{\varphi}{\bf m}({\bf r}_{i})\,,
\end{align}
from which follows
\begin{align}
\Big(\partial_{x}{\bf m}({\bf r}_{i})\Big)^2 +
\Big(\partial_{y}{\bf m}({\bf r}_{i})\Big)^2 =
\Big(\partial_{\rho}{\bf m}({\bf r}_{i})\Big)^2 + 
\frac{1}{\rho^2}\Big(\partial_{\varphi}{\bf m}({\bf r}_{i})\Big)^2.
\end{align}
Assuming that $\theta=\theta(\rho,\varphi)$ and $\psi=\psi(\rho,\varphi)$, the derivatives of the magnetization can be expressed as
\begin{align}
\partial_{\rho}{\bf m}({\bf r}_{i}) &=
\Big(\cos\theta\cos\psi\,\dot\theta_{\rho}-\sin\theta\sin\psi\,\dot\psi_{\rho}\Big)\,{\bf e}_{x} + 
\Big(\cos\theta\sin\psi\,\dot\theta_{\rho}+\sin\theta\cos\psi\,\dot\psi_{\rho}\Big)\,{\bf e}_{y} -
\sin\theta\,\dot\theta_{\rho}\,{\bf e}_{z} \,, \\
\partial_{\varphi}{\bf m}({\bf r}_{i}) &=
\Big(\cos\theta\cos\psi\,\dot\theta_{\varphi}-\sin\theta\sin\psi\,\dot\psi_{\varphi}\Big)\,{\bf e}_{x} + 
\Big(\cos\theta\sin\psi\,\dot\theta_{\varphi}+\sin\theta\cos\psi\,\dot\psi_{\varphi}\Big)\,{\bf e}_{y} -
\sin\theta\,\dot\theta_{\varphi}\,{\bf e}_{z} \,.
\end{align}
The exchange and DMI energies then equal to 
\begin{align}
E^{J}_{i} &= 
\frac{J'}{2}\left[\dot\theta^2_{\rho} + \sin^2\theta\,\dot\psi_{\rho}^2  + \frac{1}{\rho^2}\dot{\theta}^2_{\varphi} +
\frac{1}{\rho^2}\sin^2\theta\,\dot\psi^2_{\varphi} \right],\\
E^{D}_{i} &= D'\left(\cos(\psi-\varphi)\left[\dot\theta_{\rho}+\frac{1}{\rho}\sin\theta\cos\theta\,\dot\psi_{\varphi}\right] + \sin(\psi-\varphi)\left[\frac{1}{\rho}\dot\theta_{\varphi}-\sin\theta\cos\theta\,\dot\psi_{\rho}\right]\right).
\end{align}
Finally, the micromagnetic energy can be written as follows
\begin{align}
E(\theta,\psi) =\int_{0}^{\infty}{\cal E}(\theta,\psi,\rho,\varphi)\,d\rho\,d\varphi \,,
\end{align}
where the skyrmionic energy density is
\begin{align}
{\cal E}(\theta,\psi,\rho,\varphi) 
&= \frac{J'}{2}\left[\rho\dot\theta^2_{\rho} + \rho\sin^2\theta\,\dot\psi_{\rho}^2  + \frac{1}{\rho}\dot{\theta}^2_{\varphi} +
\frac{1}{\rho}\sin^2\theta\,\dot\psi^2_{\varphi} \right] -
B'\rho\cos\theta \\ 
& + D'\left(\cos(\psi-\varphi)\left[\rho\dot\theta_{\rho}+\sin\theta\cos\theta\,\dot\psi_{\varphi}\right] + \sin(\psi-\varphi)\left[\dot\theta_{\varphi}-\rho\sin\theta\cos\theta\,\dot\psi_{\rho}\right]\right). \notag
\end{align} 
The set of the Euler-Lagrange equations for this energy density
\begin{align}
\begin{cases}
\frac{\partial{\cal E}}{\partial\theta} - \frac{d}{d\rho}\frac{\partial{\cal E}}{\partial\dot\theta_{\rho}} - \frac{d}{d\varphi}\frac{\partial{\cal E}}{\partial\dot\theta_{\varphi}}=0,
\\
\frac{\partial{\cal E}}{\partial\psi} - \frac{d}{d\rho}\frac{\partial{\cal E}}{\partial\dot\psi_{\rho}} - \frac{d}{d\varphi}\frac{\partial{\cal E}}{\partial\dot\psi_{\varphi}}=0,
\end{cases}
\end{align}
then reads
\begin{align}
\left\{\hspace{-0.15cm}
\begin{matrix}
&J'\left[\rho\,\ddot\theta_{\rho} + \dot\theta_{\rho} + \frac{1}{\rho}\ddot\theta_{\varphi} - \frac{1}{\rho} \sin\theta\cos\theta\,\dot\psi^2_{\varphi} - \rho\sin\theta\cos\theta\,\dot\psi^2_{\rho} \right] +  2D'\left[\cos(\psi-\varphi)\sin^2\theta\,\dot\psi_{\varphi} - \rho\sin(\psi-\varphi)\sin^2\theta\,\dot\psi_{\rho}\right] - B'\rho\sin\theta=0 \,,\\
&J'\left[\rho\sin^2\theta\,\ddot\psi_{\rho} + \sin^2\theta\,\dot\psi_{\rho} + \rho\sin2\theta\,\dot\psi_{\rho}\dot\theta_{\rho} + \frac{1}{\rho}\sin^2\theta\,\ddot\psi_{\varphi} + \frac{1}{\rho}\sin2\theta\,\dot\theta_{\varphi}\,\dot\psi_{\varphi}\right] + 2D'\left[\rho\sin(\psi-\varphi)\sin^2\theta\,\dot\theta_{\rho} - \cos(\psi-\varphi)\sin^2\theta\,\dot\theta_{\varphi}\right] = 0 \,.
\end{matrix}
\right. \notag
\end{align}
Here we restrict ourselves to the particular case of the $C_{nv}$ symmetry. Then, one can assume that $\dot\theta_{\varphi}=0$ and $\psi-\varphi=\pi{}n$ ($n\in{\mathbb Z}$), which leads to
\begin{align}
\alpha \left[\rho^{2}\,\ddot\theta_{\rho} + \rho \dot\theta_{\rho} - \sin\theta\cos\theta \right] \pm 2\rho\sin^2\theta - \beta\rho^{2}\sin\theta=0 \,,
\end{align}
where $\alpha=J'/D'$ and $\beta=B'/D'$. Although we are interested in the problem where the exchange interaction is absent, it is still necessary to keep $\alpha\ll1$ as a small parameter in order to investigate stability of the skyrmionic solution under small perturbations. Therefore, one can look for a solution of the following form
\begin{align}
\theta = \theta_{0} + \alpha \theta_{1} + O(\alpha^2),
\end{align}
which results in
\begin{align}
\alpha \left[ \rho^{2}\,\ddot\theta_{0} + \rho \dot\theta_{0} - \sin\theta_{0}\cos\theta_{0} \right] 
\pm 2\rho\sin^2\theta_{0} \pm 4 \alpha \rho\sin\theta_{0}\cos\theta_{0}\, \theta_{1}
- \beta\rho^{2}\sin\theta_{0} - \alpha \beta\rho^{2}\cos\theta_{0}\,\theta_{1}
=0 \,.
\end{align}

\subsection{Solution for $J'=0$}
When the exchange interaction is dynamically switched off ($J' = 0$), the zeroth order in the limit $\alpha\ll1$ leads to
\begin{align}
\rho\,\sin\theta_{0} \left( \beta \rho \mp 2\sin\theta_{0} \right) =0 \,.
\end{align}
This yields two solutions:
\begin{align}
1)~\sin\theta_{0} &= 0,~\text{which corresponds to a FM ordered state}\\
2)~\sin\theta_{0} &= \pm \frac{\beta\rho}{2} = \pm \frac{B'\rho}{2D'},~\text{which describes a Skyrmion}.\label{eq:Skprofile}
\end{align} 
Then, the unit vector of the magnetization that describes a single skyrmion is equal to 
\begin{align}
{\bf m} = \sin\theta \cos(\psi-\phi)\,{\bf e}_{\rho} + \sin\theta\sin(\psi-\varphi)\,{\bf e}_{\varphi} + \cos\theta\,{\bf e}_{z} = \pm\sin\theta\,{\bf e}_{\rho} + \cos\theta\,{\bf e}_{z} = 
\frac{\beta\rho}{2}\,{\bf e}_{\rho} + \cos\theta_0\,{\bf e}_{z},
\end{align}
Importantly, the radial coordinate of the skyrmionic solution is limited by the condition $\rho\leq\frac{2D'}{B'}$. Moreover, the $z$ component of the magnetization, namely ${\bf m}_{z}=\cos\theta_0$, is not uniquely determined by the Euler-Lagrange equations. Indeed, the solution~\eqref{eq:Skprofile} with the initial condition $\theta_0(\rho=0)=\pi$ for the center of the skyrmion describes only half of the skyrmion, because the magnetization at the boundary ${\bf m}(\rho=2D'/B')$ lies in-plane along ${\bf e}_{\rho}$, which can not be continuously matched with the FM environment of a single skyrmion. Moreover, the magnetization at the larger values of $\rho$ is undefined within this solution. Therefore, one has to make some efforts to obtain the solution for the whole skyrmionic structure. 

\begin{figure}[t!]
\centering
\includegraphics[width=0.5\linewidth]{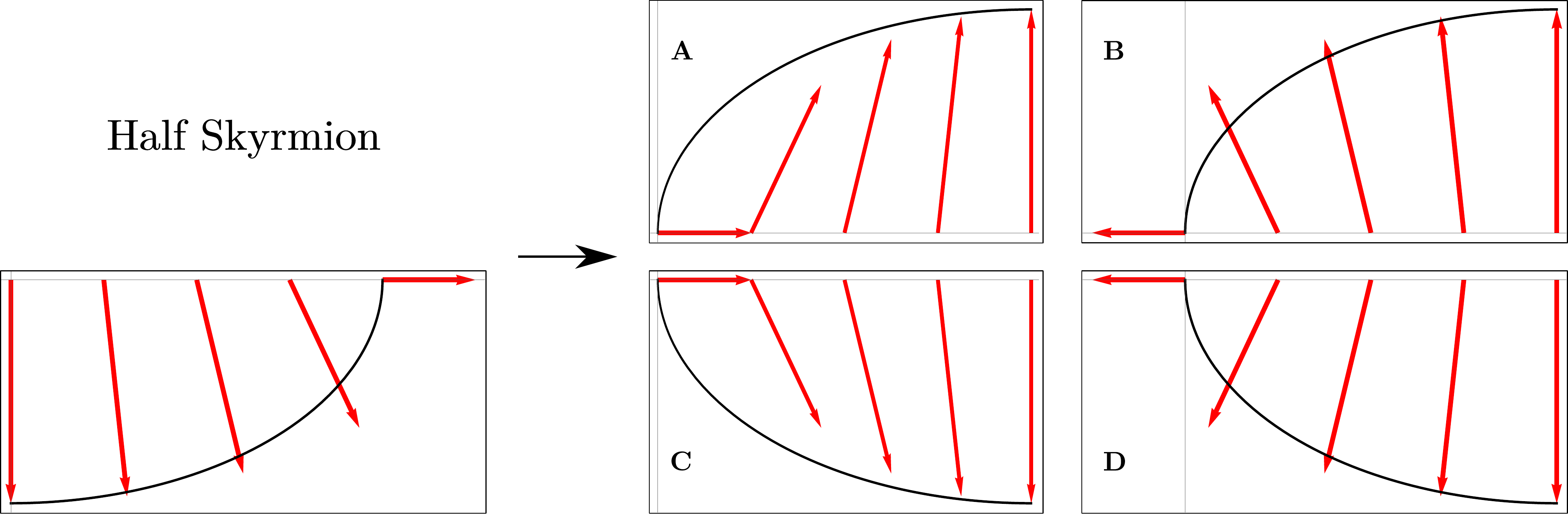}
\caption{Possible matching of the two parts of the spin spiral.}
\label{fig:SkProf}
\end{figure}

Let us stick to the case, when the magnetization of the center of the skyrmion is points, i.e. $\theta_0(\rho=0)=\pi$, ${\bf m}_{z}(\rho=0)=-1$, and magnetic field is points. Then, Eq.~\ref{eq:Skprofile} provides the solution on the segment $\theta_0\in[\pi,\frac{\pi}{2}]$ and $\rho\in[0,\frac{2D'}{B'}]$, which for every given direction with the fixed angle $\varphi$ describes the quarter period of the spin spiral as shown in the left panel of Fig.~\ref{fig:SkProf}. As it is mentioned in the main text, in the case of the $C_{nv}$ symmetry, the single skyrmion is nothing more than a superposition of three and two spin spirals for the case of the triangular and square lattice respectively. Therefore, one has to restore the second quarter of the period of a spin spiral and the rest can be obtained via the symmetry operation $\rho\to-\rho$. 
	
The second part of the spin spiral can be found by shifting the variable $\rho$ by $\rho_0$ in the skyrmionic solution~\eqref{eq:Skprofile} as $\sin\theta_0 =~ B'(\rho-~\rho_0)/2D'$. In order to match this solution with the initial one, the constant has to be equal to $\rho_0=\frac{4D'}{B'}$. Since the magnetization is defined as a continuous and differentiable function, the angle $\theta_0$ can only vary on a segment $\theta\in[\frac{\pi}{2},0]$, otherwise either the ${\bf e}_{\rho}$, or ${\bf e}_{z}$ projections of the magnetization will not fulfil the mentioned requirement. The correct matching of the two spin spirals is shown in Fig.~\ref{fig:SkProf} a), while Fig.~\ref{fig:SkProf} c) shows the violation of differentiability of ${\bf m}_{z}$ and Figs.~\ref{fig:SkProf} b), d) give a wrong matching of ${\bf m}_{\rho}$. Thus, the magnetization at the boundary of the skyrmion at $\rho=R=\frac{4D'}{B'}$ that defines the radius $R$ points up, i.e. ${\bf m}_{z}(\rho_0)=1$, which perfectly matches with the FM environment that is collinear to a constant magnetic field ${\bf B}$. 
\begin{figure}[b!]
\centering
\includegraphics[width=0.5\linewidth]{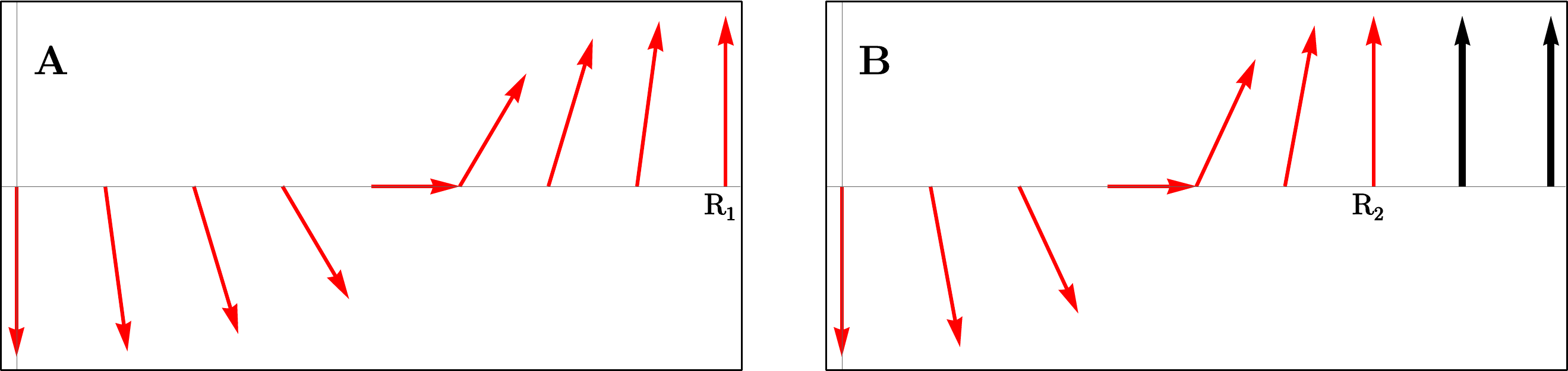}
\caption{Skyrmionic radius for two different values of the magnetic field. The larger field favours more compact structures ($R_2<R_1$) as shown in the right panel. Red arrows depict the skyrmion, while the two black arrows are related to the ferromagnetic environment.}
\label{fig:SkR}
\end{figure}
It is worth mentioning that the obtained result for the radius of the skyrmion $R=\frac{4D'}{B'} = \frac{8DSa}{B}$ is fundamentally different from the case when the skyrmion appears due to a competition between the exchange interaction and DMI. The radius in the later case is proportional to the ratio $J/D$ and does not depend on the value of the spin $S$, while in the Hiesenberg-exchange-free case it does. Although in the absence of DMI both, the exchange interaction and magnetic field, favour the collinear orientation of spins along the $z$ axis, the presence of DMI changes the picture drastically. The spins are now tilted from site to site, but the magnetic field still wants them to point in the $z$ direction and the exchange interaction aligns neighboring spins parallel without any relation to the axes. This leads to the fact that the stronger magnetic field decreases the radius of the skyrmion, while the larger value of the exchange interaction broadens the structure. This is also clear from Fig.~\ref{fig:SkR} where in the case of zero exchange interaction the larger magnetic field favours the alignment with the smaller radius of the skyrmion $R_2<R_1$ shown in the right panel.

\begin{figure}[t!]
\centering
\includegraphics[width=0.32\linewidth]{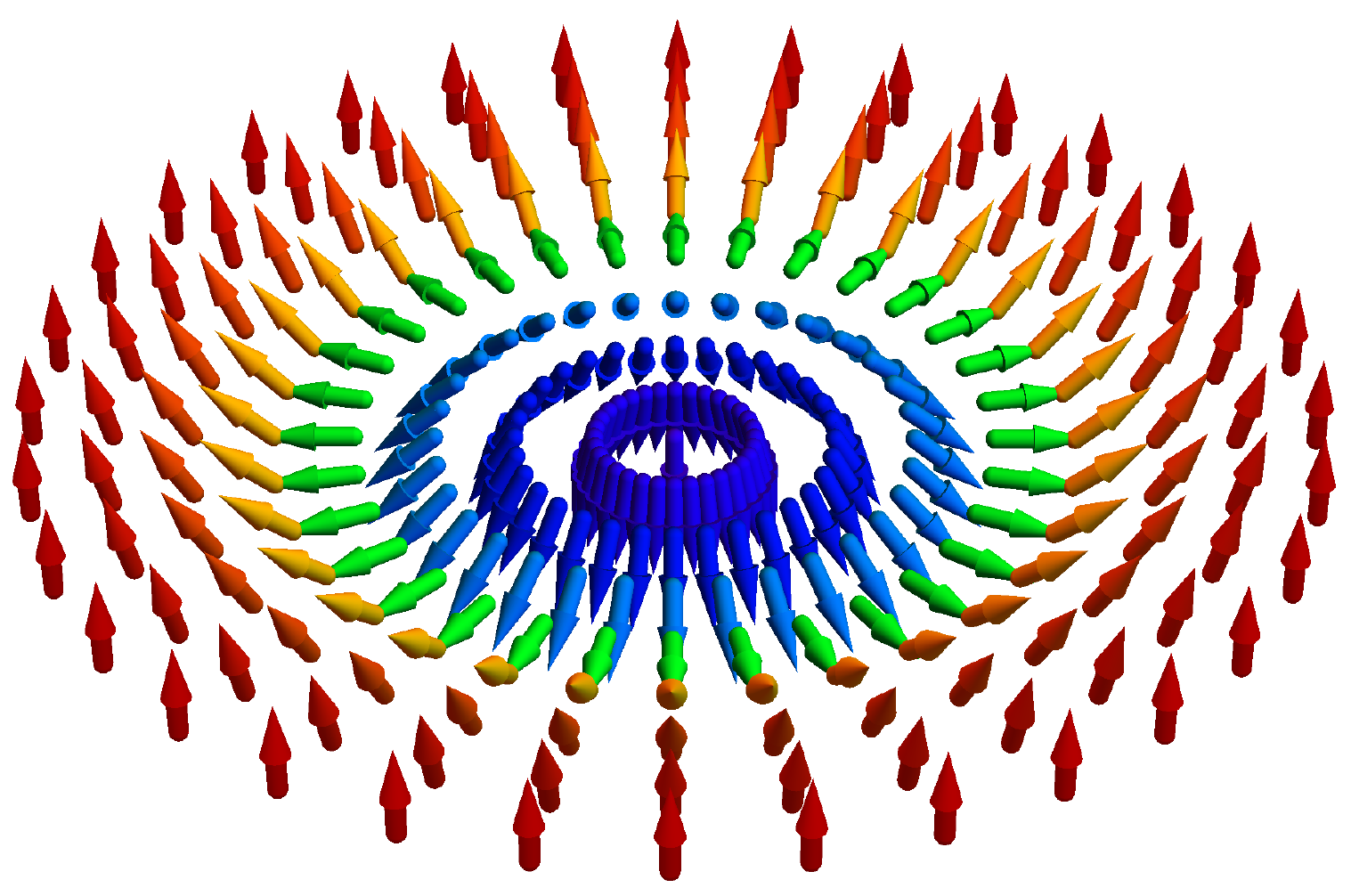}
\caption{Spatial profile of the skyrmionic solution.}
\label{fig:Sk}
\end{figure}

Finally, the obtained skyrmionic structure is shown in Fig.~\ref{fig:Sk}. It is worth mentioning that our numerical study corresponds to $B\sim{}D$, so the radius of the skyrmion is equal to $\rho_{0}\sim4Sa$, which is of the order of a few lattice sites. Although for these values of the magnetic field the micromagnetic model is not applicable, because the magnetization changes a lot from site to site, it provides a good qualitative understanding of the skyrmionic behavior and still matches with our numerical simulations.

The corresponding skyrmion number in a two-dimensional system is defined as 
\begin{align}
N=\frac{1}{4\pi} \int dx\,dy\,{\bf m} \left[ \partial_{x}{\bf m}\times\partial_{y}{\bf m}\right]
\end{align}
and then equal to 
\begin{align}
N=\frac{1}{4\pi} \int dx\,dy\,\frac{1}{\rho}\sin\theta\,\dot\theta_{\rho} = \frac{1}{4\pi} \int d\rho\,d\varphi\,\sin\theta\,\dot\theta_{\rho} = \frac12\left(\cos\theta(0) - \cos\theta(\rho_{0})\right) = 1.
\end{align}
One can also consider the case of zero magnetic field. Then, solution of the Euler-Lagrange equations
\begin{align}
\left\{
\begin{matrix}
\dot\psi_{\varphi} = \rho\tan(\psi-\varphi)\,\dot\psi_{\rho},\\
\dot\theta_{\varphi} = \rho\tan(\psi-\varphi)\,\dot\theta_{\rho}
\end{matrix}
\right.
\end{align}
describes a spiral state, as shown in Fig.~\ref{spinfactors}.

\subsection{Solution for small $J'$}
Now, let us study stability of the skyrmionic solution \eqref{eq:Skprofile} and consider the case of a small exchange interaction with respect to DMI. The first order in the limit $\alpha\ll1$ implies
\begin{align}
\left[ \rho^{2}\,\ddot\theta_{0} + \rho \dot\theta_{0} - \sin\theta_{0}\cos\theta_{0} \right]
\pm 4 \rho\sin\theta_{0}\cos\theta_{0}\, \theta_{1}
- \beta\rho^{2}\cos\theta_{0}\,\theta_{1}
=0 \,.
\end{align}
The zeroth order solution leads to
\begin{align}
\cos \theta_{0} \, \dot\theta_{0} = \pm \frac{\beta}{2} ~~~\text{and}~~~ \cos\theta_{0} \, \ddot\theta_{0} -\sin\theta_{0} \, \dot\theta_{0}^{2} = 0 \,.
\end{align}
This results in
\begin{align}
&\rho^{2}\frac{\sin\theta_{0}}{\cos\theta_{0}} \left(\frac{\beta}{2\cos\theta_{0}}\right)^{2} \pm \rho \frac{\beta}{2\cos\theta_{0}} - \sin\theta_{0}\cos\theta_{0} = - \beta \rho^{2} \cos\theta_{0} \, \theta_{1}
\notag\\
&\pm \left(\frac{\beta\rho}{2}\right)^{3}\frac{1}{\cos^{4}\theta_{0}} \pm \frac{\beta\rho}{2} \frac{1}{\cos^{2}\theta_{0}} \mp \frac{\beta\rho}{2} = - \beta \rho^{2} \, \theta_{1}
\notag\\
&\beta \rho^{2} \, \theta_{1} = \mp \left[ \left(\frac{\beta}{2} \rho \right)^{3} \frac{1}{\cos^{4}\theta_{0}}  + \frac{\beta}{2} \rho \left( \frac{1}{\cos^{2}\theta_{0}} - 1 \right) \right] \notag\\ 
&\theta_{1} = - \frac{\beta}{4}\sin\theta_{0} \left[ \frac{1}{\cos^{4}\theta_{0}} + \frac{1}{\cos^{2}\theta_{0}} \right]
\,,
\end{align}
provided $\cos\theta_{0}\neq0$. Therefore the total solution for the skyrmion 
\begin{align}
\theta = \theta_{0} - \frac{J'B'}{4D'^2}\sin\theta_{0} \left[ \frac{1}{\cos^{4}\theta_{0}} + \frac{1}{\cos^{2}\theta_{0}} \right]
\end{align}
is stable in the two important regions when $\sin\theta_0=0$ -- around the center of skyrmion and at the border. The divergency of the correction $\theta_1$ in the middle of the skyrmion when $\cos\theta_0=0$ comes from the fact that the magnetization is poorly defined here, as it was discussed above.


\begin{thebibliography}{42}
\bibitem{ref}
Although the rich variety of nanoskyrmions on the discrete lattices obtained in the current study can not be described by the micromagnetic model, because the length scale on which the magnetic structure varies is of the order of the interatomic distance, the result for the radius matches very well our numerical simulations and can be considered as a limiting case of the micromagnetic solution when the applied magnetic filed is of the order of DMI. Nevertheless, the analytical solution for the isolated skyrmion might be helpful for the other set of system parameters, where the considered micromagnetic model is applicable.
\end{thebibliography}
\end{document}